\documentclass[journal]{IEEEtran}
\usepackage{cite}
\usepackage{amsmath,amssymb,amsfonts}
\usepackage{graphicx}
\usepackage{enumerate}  
\usepackage{textcomp}
\usepackage[per-mode=symbol,per-symbol=/]{siunitx}
\usepackage{hyperref}
\DeclareSIUnit\bit{b}
\usepackage{enumerate}  
\usepackage{subcaption}
\usepackage{makecell}
\usepackage{adjustbox}
\usepackage{booktabs}
\usepackage{url}

\begin{document}
%
\title{
\footnotesize This work has been submitted to the IEEE for possible publication.\\Copyright may be transferred without notice, after which this version may no longer be accessible.\\
\Huge
Towards Specialized Wireless Networks Using an ML-Driven Radio Interface
}

%
%
%

\author{Kamil Szczech, 
        Maksymilian Wojnar,
        Katarzyna Kosek-Szott, 
        Krzysztof Rusek, 
        Szymon Szott, 
        Dileepa Marasinghe, 
        Nandana Rajatheva, 
        Richard Combes, 
        Francesc Wilhelmi, 
        Anders Jonsson, 
        and~Boris Bellalta
\thanks{Kamil Szczech, 
        Maksymilian Wojnar,
        Katarzyna Kosek-Szott, 
        Krzysztof Rusek, 
        and Szymon Szott are with the AGH University of Krakow, Poland,
        e-mail: szymon.szott\@agh.edu.pl}
\thanks{Dileepa Marasinghe and Nandana Rajatheva are with University of Oulu, Finland.}
\thanks{Richard Combes is with CentraleSupélec, France.}
\thanks{Francesc Wilhelmi, 
        Anders Jonsson, 
        and~Boris Bellalta are with Universitat Pompeu Fabra, Barcelona, Spain.}
}

\maketitle

\begin{abstract}
Future wireless networks will need to support diverse applications (such as extended  reality), scenarios (such as fully automated industries), and technological advances (such as terahertz communications). Current wireless networks are designed to perform adequately across multiple scenarios so they lack the adaptability needed for specific use cases. Therefore, meeting the stringent requirements of next-generation applications incorporating technology advances and operating in novel scenarios will necessitate wireless specialized networks which we refer to as SpecNets. These networks, equipped with cognitive capabilities, dynamically adapt to the unique demands of each application, e.g., by automatically selecting and configuring network mechanisms. An enabler of SpecNets are the recent advances in artificial intelligence and machine learning (AI/ML), which allow to continuously learn and react to changing requirements and scenarios. By integrating AI/ML functionalities, SpecNets will fully leverage the concept of AI/ML-defined radios (MLDRs) that are able to autonomously establish their own communication protocols by acquiring contextual information and dynamically adapting to it. In this paper, we introduce SpecNets and explain how MLDR interfaces enable this concept. We present three illustrative use cases for wireless local area networks (WLANs): bespoke industrial networks, traffic-aware robust THz links, and coexisting networks. Finally, we showcase SpecNets' benefits in the industrial use case by introducing a lightweight, fast-converging ML agent based on multi-armed bandits (MABs). This agent dynamically optimizes channel access to meet varying performance needs: high throughput, low delay, or fair access. Results demonstrate significant gains over IEEE 802.11, highlighting the system’s autonomous adaptability across diverse scenarios.
\end{abstract}


%
\IEEEpeerreviewmaketitle

\section{Introduction}
\IEEEPARstart{W}{ireless} networks are a cornerstone of our fully connected digital society. By enabling connectivity for people, machines, robots, and other objects with communication interfaces, wireless networks not only transform how we work, produce, learn, and entertain but also influence the trajectory of human evolution. In the coming years, we will witness the widespread adoption of extended reality (XR) and virtual reality (VR) technologies, the advancement of the Internet of things (IoT) alongside autonomous robots (terrestrial and aerial), fully automated industries, smart manufacturing plants, and intelligent infrastructures and environments, to name just a few emerging applications and technological breakthroughs~\cite{giordani2020toward}.

To support the diverse applications, scenarios, and technological advances outlined above, traditional wireless approaches---relying on general-purpose solutions with incremental updates to address varying needs---prove costly and inefficient. These systems often incorporate complex technologies and mechanisms governed by static parameters, rules, and protocols. Although designed to perform adequately across multiple scenarios, they lack the adaptability needed to fully optimize performance for specific use cases. Meeting the stringent requirements of next-generation applications in throughput, latency, reliability, connectivity, and power efficiency will necessitate wireless specialized networks (SpecNets). These networks, equipped with cognitive capabilities, can dynamically adapt to the unique demands of each application, e.g., by allocating appropriate resources or automatically selecting and configuring network mechanisms. This paper explores SpecNets as a critical evolution beyond general-purpose networks, enabling devices to autonomously (i.e., without the need for highly specialized and trained staff to manually adapt network configurations) tailor their behavior to meet the exacting needs of emerging applications.

\begin{figure*}[!t]
\centering
\includegraphics[width=0.85\linewidth]{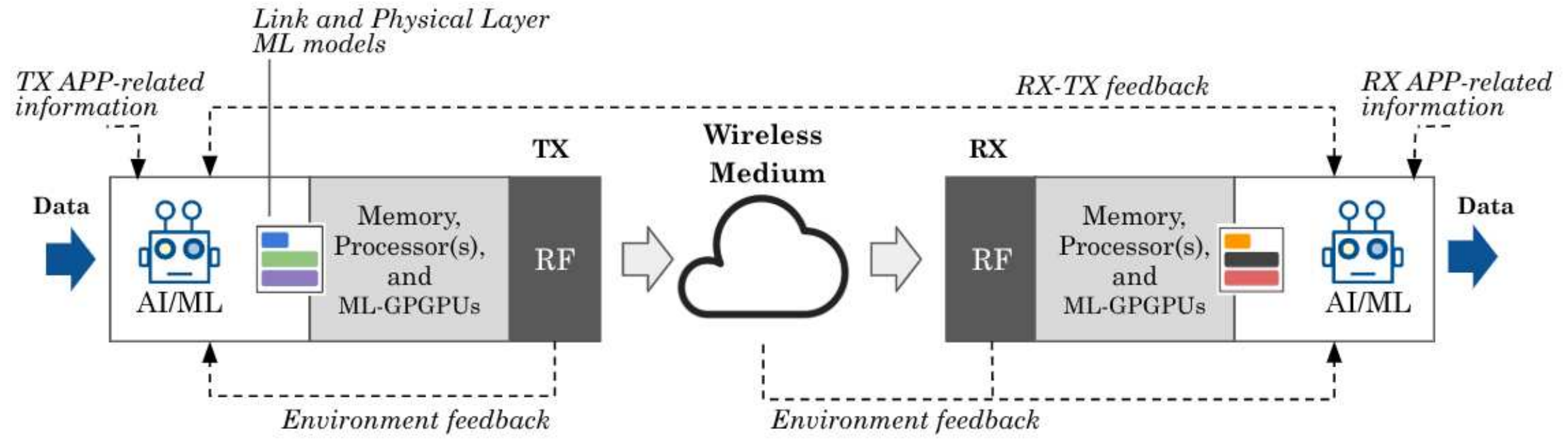}
    \caption{Conceptual MLDR architecture \cite{bellalta2024towards}}
    \label{fig:MLDR_architecture}
\end{figure*}

A critical enabler for the development of SpecNets are the recent advances in artificial intelligence and machine learning (AI/ML) which allow to process data differently than current approaches by extracting useful patterns from complex data. This new information (e.g., based on new processed information coming from the network or even from other sources such as cameras or sensors) provides networks with more sophisticated capabilities such as cognition.
Indeed, artificial intelligence mirrors the `human cognitive cycle', comprising stages of data acquisition, knowledge extraction, reasoning, and action, ultimately establishing causal relationships between actions and outcomes. Within this cycle, ML models play a pivotal role by extracting patterns and insights from data and encapsulating them in a ``learned from experience'' function. Therefore, AI/ML solutions offer the necessary capabilities to allow SpecNets to continuously learn and react to changing requirements and scenarios in real-time. By integrating AI/ML functionalities, as outlined in~\cite{bellalta2024towards,wilhelmi2024machine}, SpecNets will fully leverage the concept of AI/ML-defined radios (MLDRs) that are able to autonomously establish their own communication protocols by acquiring contextual information and dynamically adapting to it.

In this paper, our main contribution is the introduction of wireless SpecNets and explaining how MLDR interfaces enable this concept. 
After a review of related research on AI-native specialized wireless networks (Section~\ref{sec:ai_native_protocols}), we next present three illustrative use cases of SpecNets in Section~\ref{sec:usecases}:
\begin{enumerate}[(i)]
    \item bespoke industrial networks,
    \item traffic-aware robust THz links,
    \item coexisting networks.
\end{enumerate}
For each use case, we show how MLDRs exploit their learning abilities to meet the unique challenges and relevant requirements. 
Next, we provide an initial evaluation of the potential of SpecNets to improve network performance.
Although MLDR-based SpecNets are a universal concept that applies to any type of technology, we showcase the potential of our solution within the domain of IEEE 802.11 technology (aka Wi-Fi).
Specifically, in Section~\ref{sec:results}, we consider industrial networks and show a lightweight, rapidly converging ML agent based on multi-armed bandits (MABs).
The agent dynamically learns and drives access to the channel to meet different performance needs (high throughput, low delay, or fairness in channel access), i.e., the same agent can achieve diverse goals. The implementation and simulations are conducted with \mbox{ns-3}, using ns3-ai~\cite{ns3ai} to establish a communication interface between the simulator and the ML agent. The results reveal significant improvements in both latency and throughput compared to current IEEE 802.11, showing the system's autonomous adaptability across diverse requirements and scenarios.
Finally, we conclude the paper and outline future work in Section~\ref{sec:conclusions}.

\section{AI/ML-defined Radios (MLDRs)}

An MLDR natively supports AI/ML-based optimization and decision-making in its communication functions (such as data transmission and reception, channel access and frame exchange, and radio resource management). With such features, an AI/ML-defined radio is expected to improve the efficiency of any required communication (e.g., exchange of AI metadata among different network devices), reduce existing overheads, and make better use of resources. In addition, such an interface will enable more efficient transmitter and receiver implementations and higher flexibility in diverse and changing scenarios. 

From an architectural standpoint, MLDRs diverge from traditional systems that are optimized for specific and predefined firmwares. MLDRs should be designed around or at least incorporate ML accelerators to enable rapid ML model inference. Additionally, if on-device training is required, additional storage and computational resources are needed to avoid compromising transmit/receive (TX/RX) functionalities. Finally, regarding the design of MLDRs, although fully flexible architectures theoretically offer the advantage of supporting continuously evolving AI/ML models, it is more probable that we will see circuits optimized to efficiently execute only specific types of ML models. However, this specialization may limit the ability of the radio to quickly adapt to emerging ML models. Figure~\ref{fig:MLDR_architecture} illustrates the modular concept of an MLDR, wherein ML agents and models govern the radio processing resources and radio frequency (RF) hardware, the frame exchange between the transmitter and the receiver, and the signals exchanged over the air.

Before MLDRs become a reality, performance gains must be consistently observed. In particular, MLDRs should also be evaluated considering their adaptability to different situations, which means they will have to provide the required service levels in many more situations than nowadays, with few to no exceptions. Therefore, to evaluate an AI/ML-driven radio, in addition to raw performance, we have to consider aspects such as training time and generalization capabilities to different situations. 

MLDRs need to perform regular model exchanges, deployment, and retraining to update the ML models in place. This, in turn, requires additional infrastructure and processes to be held both in the radio itself (e.g., low-cost model inference) and in cloud/edge equipment (e.g., data storage, model training, distributed learning orchestration). Moreover, considering that data is the fuel for applying ML, a huge amount of ML data, including not only training data but also ML model data and metadata, must be exchanged throughout different parts of the network. 


In this paper, we envision that the configuration of an MLDR is governed by an agent, which implements logical functionalities for running dynamically instantiated AI operations, according to the required pipeline for each use case and scenario; in a real network, the ML part could be hosted at the AP or a centralized network controller communicating with the AP. In particular, the agent must be able to update its state based on network performance metrics, such as aggregate throughput or mean latency, thereby requiring to collect network measurements. 
Then, the agent might select the network parameter values and send them back to the MLDR, adhering to the standard reinforcement learning (RL) framework. The retrieval of network measurements and exchange of commands can occur periodically, within intervals that we refer to as \textit{control periods}. Apart from that, to support the interaction with other agents, a coexistence module is envisioned, which would allow, for instance, running distributed ML optimization (e.g., federated learning).

\section{AI-native Wireless Networking: Current Views}
\label{sec:ai_native_protocols}

In this section, we summarize current views to realize AI-native networking by enabling AI/ML-defined radios, focusing either on its conceptual approach or individual functionalities. For the individual functionalities, we focus on link-layer protocols, and the use of ML agents able to build them. Finally, we provide some input on the actual ML strategies able to power them.


\subsection{A Conceptual Approach}

The use of ML in the realm of communications has greatly progressed in the last few years, and the vision of how to realize such an integration has continuously evolved since then. Initial efforts in industry and standardization led to on-top ML implementations, where ML operations were designed to support the existing network infrastructure and functionalities~\cite{wilhelmi2020flexible, szott2021wifi}. Such an integration was conceived as ML overlays interacting with network components (e.g., to get information or give commands) through existing or slightly new high-level interfaces.
Today, AI-native solutions are being envisioned to bring ML operations as an intrinsic component of network operations and standards~\cite{lin2024overview, hoydis2021toward}. 
The Open Radio Access Network (O-RAN) interoperability initiative for cellular networks also considers AI/ML workflows \cite{polese2023understanding}, although this approach is rather an example of AI/ML-driven control than AI/ML-radio. 
Moreover, the sudden irruption of Generative AI (GAI) and large language models (LLMs) 
has redefined the way of thinking in the sector~\cite{maatouk2024large}.


An exemplary conceptual approach is proposed in \cite{valcarce2024role}, where the authors describe a native AI-based air interface envisioned for 6G. Their primary focus is on AI-driven MAC functions, including protocol learning, which enables tailoring data link layer signaling to specific network deployments and needs (e.g., reduced signaling overhead and energy consumption in the case of sporadic transmissions in sensor networks).
Furthermore, Park et al.~\cite{park2023towards}  categorize ML-driven MAC protocols into three levels. 
Level 1 comprises multi-agent deep reinforcement learning (DRL)-based protocols, which are task-oriented but consider task knowledge as a black box. 
Level 2 comprises neural network-oriented symbolic protocols \cite{seo2023toward}, which provide interpretable MAC operation and allow direct protocol adjustment. 
Finally, level 3 comprises language-oriented semantic protocols using both large language and generative models. 
The authors conclude that each approach has its disadvantages. 
Level 1 protocols struggle with computing efficiency (over-parameterized neural networks, high computation/communication/memory costs), management flexibility, generalization, black-box-like operation. 
Level 2 protocols are not well suited for non-stationary task environments and changing requirements since in such cases they need neural network re-training. 
Finally, level 3 protocols tend to have large computation overheads and experience high latency during execution. 

As shown above, the conceptual design of AI-native wireless networking is still in its infancy and requires additional considerations. 
However, some important challenges on the path towards MLDRs have already been identified and are gradually being addressed by the researchers. 
In the following, we present state-of-the-art AI-native wireless link-layer protocols as examples.

\subsection{AI-native MAC protocols}

Several approaches to building and implementing AI-native wireless MAC protocols have been proposed in the literature. Most typically, the authors propose selecting ML agent actions based on a predefined set of MAC functions or modules (e.g., select the value of a parameter among a predefined list of possible values) in order to build customized MAC protocols tailored to particular network scenarios. 

Keshtiarast et al. \cite{keshtiarast2024ml} propose an ML framework for wireless MAC protocols tailored to individual application needs.\footnote{Publicly available implementation:  \url{https://gitlab.com/navid-keshtiarast/Rec\_mac}.} They implement a proximity policy optimization algorithm (a type of DRL following the actor-critic approach), which enables adding, removing, and modifying protocol features based on interactions with the network environment. The authors assume that the MAC policy is shared among all nodes. The agent adjusts its behavior by selecting actions based on an array of IEEE 802.11ac MAC functions (e.g., carrier sensing, back-off, request to send and clear to send (RTS/CTS) mechanism) and their parameters (e.g., back-off type, slot time length) considering environment characteristics (e.g., traffic patterns, number of stations). The proposed method is shown to obtain higher throughput and lower latency compared to the legacy 802.11ac MAC protocol in static deployments. However, some of the assumptions seem unrealistic, e.g., the fully observable environment. 

In \cite{pasandi2021towards}, the authors propose a general framework based on DRL for the design and evaluation of network protocols. As an example, the 802.11 MAC protocol is divided into a set of parametric modules (RTS/CTS, back-off, fragmentation) which then serve as an input to the DRL agent. The resulting DeepMAC mechanism was shown to outperform legacy distributed coordination function (DCF) in terms of throughput. 
Additionally, the authors have noticed several challenges in building customized MAC protocols, among which, unreliable MAC protocol performance seems most crucial. In DeepMAC design, the authors assume that the acknowledgments are handled by upper-layer protocols, which, however, may negatively impact, e.g., the performance of TCP, which could mistaken poor channel conditions with network congestion.

A partly similar RL-based approach is proposed in \cite{pan2024mac}. A new AI-MAC employs ML to adapt to changing network conditions, avoid interference, optimize channel access, and provide deterministic delay in Wi-Fi networks. The proposed framework is based on the centralized training and decentralized execution paradigm. Each AI-MAC station contains two ML agents responsible for rate control and channel access. Both agents are orchestrated by a global device agent and work asynchronously. In addition, QoS requirements are monitored to optimize AI-MAC operation for real-time traffic. AI-MAC was tested against 802.11ax and 802.11be in three different scenarios (home, office, shopping mall) and was shown to obtain lower jitter and latency in each scenario.  


The presented examples constitute an important step towards MLDRs and AI-native MAC protocol design. 
They provide useful insights for future developments and confirm that MLDR reasoning and learning capabilities can be obtained with the use of AI/ML agents to support the required SpecNets capabilities.  
Importantly, in case of dynamic SpecNets environments, the collaboration of multiple agents will be required, as we discuss next.


\begin{table*}[htb]
    \centering
    \caption{Key features and KPIs of the three main SpecNets services and corresponding RL learning objectives.}
    \label{tab:services}    
    \begin{tabular}{@{}lllll@{}}
        \toprule
        \textbf{SpecNet service}                        & \textbf{No. of users} & \textbf{Throughput} & \textbf{Latency} & \textbf{RL objective}                      \\ \midrule
        High throughput & 1-50                  & Maximum              & No requirements        & Maximize throughput                               \\
        Low latency  & 1-10                 & Low                 & Below threshold      & Minimize latency 
        \\
        Massive communication           & \textgreater{}100    & Low                   & No requirements        & Maximize channel access fairness
        \\ \bottomrule
    \end{tabular}
\end{table*}

\subsection{Multi-agent Learning Systems}

A defining characteristic of wireless networking is the need for distributed, multi-agent learning algorithms. Multi-player bandit models~\cite{kalathil2014decentralized} are a particularly relevant tool for analyzing scenarios in which agents must collaboratively identify optimal transmission strategies (e.g., distributed channel access) while optimizing performance. Recent advancements~\cite{rosenski2016multi,besson2018multi} propose algorithms tailored to different levels of agent interaction---whether information exchange is permitted and whether agents can dynamically join or leave the system---offering performance guarantees. However, even in basic settings, optimal algorithms remain elusive~\cite{huang2022towards}, and it is unclear if distributed learning can achieve the same convergence speed as centralized learning, especially for competitive settings where agents are set to learn selfishly~\cite{wilhelmi2019collaborative}. When agents are able to exchange information, MLDR interfaces can utilize joint learning strategies, such as federated learning, to enhance collaboration.

Multi-agent learning problems in wireless networking often feature inherent structures (e.g., combinatorial structures for resource allocation~\cite{combes2015combinatorial}, uni-modal structures for link adaptation and rate selection~\cite{combes2018optimal}, and graphical structures for routing~\cite{talebi2017stochastic}) that must be exploited to derive feasible solutions. Furthermore, these problems involve navigating a trade-off between learning speed and computational complexity~\cite{cuvelier2021statistically}, as nodes may have limited processing power and time for decision-making. Bridging the gap between theoretical insights and practical implementations requires continued research to develop more efficient algorithms, robust performance evaluations, and scalable architectures for AI-driven future communication networks.

The inherent structures described above imply that several processes have to be optimized separately during operation. In this setting, we foresee that hierarchical RL will play an important role in the trade-off between learning speed and computational complexity. In hierarchical RL, a learning agent can decide which process to optimize next, and the agent then focuses exclusively on this process until a given termination condition is satisfied. During optimization of an individual process, the agent can ignore all the features associated with other structures, significantly simplifying learning.

\section{Use cases: Requirements, technologies, and KPIs}
\label{sec:usecases}

In this section, we discuss the three use cases that showcase the usability of SpecNets: bespoke industrial networks, robust high-speed (THz) links, and competing networks. Each use case shows different areas where the application of MLDRs provides performance gains.

\subsection{Bespoke Industrial Networks}
\label{sec:bespoke}

\subsubsection{Motivation}

\begin{figure}[!t]
\centering
\includegraphics[width=0.95\columnwidth]{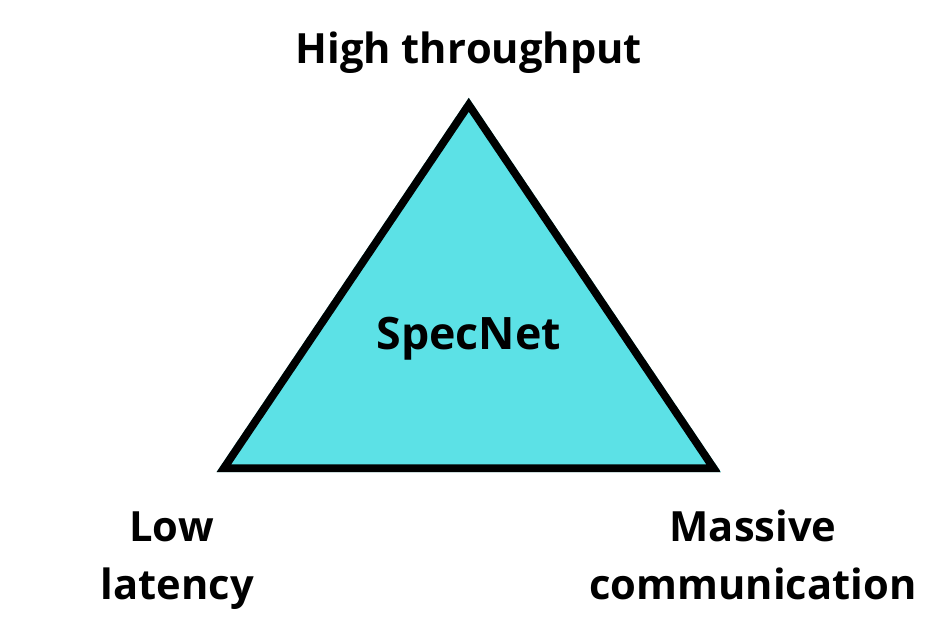}
    \caption{The SpecNet triangle for bespoke industrial networks. Operators specify their requirements as a point within this triangle, thereby distributing the importance of each of the three conflicting requirements.}
    \label{fig:triangle} 
\end{figure}

Stand-alone wireless industrial networks are typically task-oriented, where the emphasis is on the optimization of a specific goal \cite{seferagic2020survey}. 
Therefore, such networks should be flexible in their configuration to meet the needs of the main applications, hence the term bespoke networking.
The underlying goal can be formulated, e.g., as achieving high throughput, low latency, or assuring channel access fairness.\footnote{These are only examples and other metrics can be considered, e.g.,  a low latency percentile could be also indicative of reliability.} Each of these goals requires a different set of functionalities and parameters from the network. In addition, a bespoke network's specific goals may evolve over time, and both the environment and the network themselves may vary quickly (e.g., topology changes, equipment upgrades, mobility of the machinery). 
Therefore, here the task of SpecNets is to adapt the network to operator requirements, i.e., following the principles of intent-based networking \cite{leivadeas2022survey}. 
The arising simplicity and flexibility of network configuration could significantly reduce maintenance costs and open the way for innovative industry applications.

\subsubsection{Description}

We assume that the operator of an industrial network formulates their requirements as a point on the ``SpecNet triangle'' (Figure~\ref{fig:triangle}). 
Table~\ref{tab:services} shows example requirements for the extreme corners of the triangle (although these requirements can change depending on the technology in place and the type of deployment). 
These performance limits can vary over time (due to changes in traffic, utilization patterns, changes in topology, equipment, etc.) and such changes should be automatically captured. Here, AI/ML could be again be used, leading to new ideas.

\begin{figure}[!t]
\centering
\includegraphics[width=0.8\columnwidth]{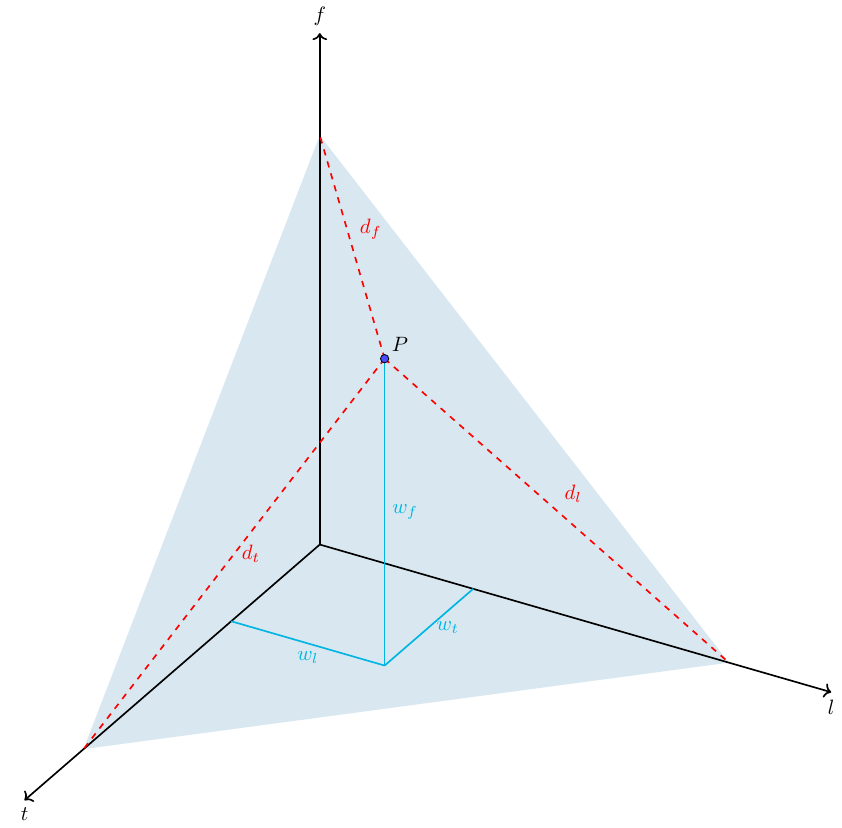}
    \caption{Unit simplex as a triangle in the SpecNets objectives' space.}
    \label{fig:3d_triangle}
\end{figure}

Given the multi-objective nature of the problem, the network operator specifies the preferences for the three objectives (high throughput, low latency, massive communications). Then, by mapping specific preferences into SpecNet triangle points (which correspond to a bijection of the Pareto front), one can easily visualize the objective space (Figure~\ref{fig:3d_triangle}).
Then, these preferences are communicated to a controller running the network's SpecNet agent, which needs to interpret the preference and translate it into achieving network performance metrics.
For numerical optimization, assume that the triangle is a simplex in the objective space where each corner corresponds to a single objective. 
Then, mixed preferences correspond to the points inside the triangle.
Formally, the triangle is a unit simplex in the objectives' space.
The distances to the corners determine the weight of the particular objective, according to the following implicit formula
\begin{equation}
    \begin{array}{c}
 (w_t-1)^2+w_l^2+w_f^2=d_t^2 \\
 w_t^2+(w_l-1)^2+w_f^2=d_l^2 \\
 w_t^2+w_l^2+(w_f-1)^2=d_f^2 \\
 w_t+w_l+w_f=1, \\
\end{array}
\end{equation}
where $d_t$, $d_l$ and $d_f$ are the distances  mapped to the exact coordinates $w_t,w_l,w_f$.
This formula guarantees that the weights are positive, they add up to one and every weight is in the range $[0,1]$.
Furthermore, this formulation can be the foundation of multi-criteria optimization.
This proposal aims to keep the perception of distance as a weighting factor, while forcing the points to be inside the simplex.
This makes the 3D coordinates of the points a proper set of weights for their linear scalarization, which is the most popular scalarization
function~\cite{drugan2013moma} for combing multiple criteria into a single numerical value.
Here we rely on the fact that there are  thre aspects of SpecNet. Given the constraint that the set of predefined weights must add up to one, we have only two degrees of freedom that can be visualized on the plane as triangle.
Network preferences can be specified similarly to color specification on a color triangle~\cite{netravali1995digital}.

\subsubsection{Related Work}
Task-oriented networks, i.e., machine-type communication and industrial IoT (IIoT), are studied in \cite{mohammed2019mission, zhou2021machine, salau2022recent}. In these examples, the task-oriented networks provide support for different kinds of mission-critical applications, e.g., public safety, industrial automation, asset tracking, healthcare, and robotics. Achieving these goals poses a significant challenge from the wireless networking perspective, due to the networks' stringent requirements for high reliability, low latency, and availability, as well as the support of a massive number of ubiquitous connections.

The adoption of IEEE 802.11 for industrial delay-sensitive wireless control and monitoring applications is surveyed in \cite{cheng2019adopting}. The authors show that contention-based protocols are easy to implement and operate satisfactorily in case of a light network load. However, contention-free protocols (\mbox{TDMA-,} polling-, and token-based) may be required to support delay-sensitive applications. Unfortunately, such protocols are harder to implement, especially in heterogeneous environments. 
Meanwhile, Wi-Fi is evolving towards reliability thanks to mechanisms such as OFDMA, TWT, shared TXOP, and multi-AP coordination \cite{galati2024what}.
Independently, some research papers resort to ML to meet the goals of task-oriented networking. 
Aboubakar et al. \cite{aboubakar2020efficient} use different ML algorithms (k-nearest neighbors, random forest, decision tree, multilayer perceptron) to optimize IEEE 802.15.4 MAC parameter settings according to network characteristics and traffic state to improve end-to-end delay.
Savaglio et al. \cite{savaglio2019lightweight} propose an energy-preserving MAC protocol for wireless sensor networks. The proposed solution is based on light Q-learning, which adjusts the duty cycle of the sleep and awake periods of the sensor nodes so that energy is saved without worsening the packed delivery ratio.
Kherbache et al. \cite{kherbache2022reinforcement} survey RL-based TDMA scheduling in IIoT, which improves adaptability and efficiency compared to conventional scheduling mechanisms. In the surveyed articles, single-channel and multi-channel approaches are analyzed, which typically concentrate either on minimizing energy consumption (e.g., wasted in case of collisions) while ensuring satisfactory throughput or delay-sensitive communications. The most commonly used RL methods are Q-learning and DRL. 
Apart from ML, deterministic MAC operation is also proposed for industrial sensor networks, e.g., Farag et al. \cite{farag2018delay} propose stealing time slots assigned to periodic non-critical traffic by aperiodic time-critical traffic. Raza et al. \cite{raza2018novel} propose a similar approach with a central controller. In case of asynchronous emergency communications, regular transmissions are stopped or a transmission opportunity on a separate control channel is requested.

In summary, the presented articles show that bespoke industrial networks require dedicated MAC solutions, which can be supported by ML. These papers also confirm that different types of networks have varying quality of service requirements. Therefore, it is required a general solution that adapts to the currently addressed task in a specific environment.  

\subsection{Data-driven Waveforms for THz Links}

\subsubsection{Motivation}

Next-generation wireless XR applications will demand high throughput (e.g., hundreds of Mbps), ultra-low latency (under 1 ms), and exceptional reliability (deterministic delay with minimal packet loss). Meeting these stringent requirements calls for dynamic adjustments in both link and physical layer protocols, aligning them with the specific traffic characteristics, application performance demands, and environmental conditions. On the technology side, sub-THz and THz bands are particularly promising for these specific tasks. Higher frequencies in the sub-THz/THz range offer high bandwidth, enabling high-throughput communication links, so higher reliability can be achieved by employing more redundancy with abundant bandwidth. However, challenges include high path loss and limited dispersion, requiring precise directional beam-forming for robust links, while at the link level, increasing hardware impairments due to high frequency and wider bandwidth, requires meticulous physical layer design. 

Therefore, a key challenge in establishing robust THz links is resilient waveform design against hardware impairments, which are more pronounced in these frequencies. Increased phase noise, which arises when synthesizing THz frequencies by up-conversion using practical oscillators and inefficient power amplifiers, are two main impairments. Utilizing AI/ML methods to tackle these challenges enables the use of data-driven, model-based techniques. Furthermore, the use of a data-aided design allows for the use of more realistic models for impairments rather than relying on less accurate models that simplify the problems for the sake of tractability and block-based optimization in conventional designs. Although the ML-driven solutions offer significant performance gains, the dynamic selection of solutions for varying demands of different tasks remains challenging since these solutions are specialized for the trained conditions, usually representing a subset of operating ranges for a task, hence lacking generalization power.

\subsubsection{Description}
The main aim of employing the sub-THz band is to achieve higher throughput by utilizing a wider bandwidth while ensuring reliability under varying requirements and conditions. In the MLDR approach, this is tackled in three fronts. First, to ensure the performance of data-driven waveforms across a wide range of conditions, generalization aspects will be explored. For example, generalizability can be incorporated in utilized hardware models during the design/training phase of the waveforms. Next, a method for dynamic selection of the data-driven waveforms based on the requirements and conditions is explored. Finally, to guarantee performance using ML-driven solutions, a monitoring mechanism that observes the performance to trigger a reselection of a better waveform when required or even a retraining phase, if possible, will be explored.
In terms of KPIs, the main KPI is the achievable data rate ($R$) with a given bandwidth. To ensure reliability, the block error rate (BLER) needs to be quantified, while spectral efficiency (SE), as defined in \eqref{eq:spec_eff}, is another KPI worth looking at. The SE is calculated as,
\begin{equation}  \label{eq:spec_eff}
    {\text{SE}}= \frac{(1-\text{BLER}) R}{\text{BW}_{\rm eff}}
\end{equation}
and $R$ is the data rate given by,
\begin{equation}
    R =  \frac{r M \rho}{T_s}
\end{equation}
where $r$, is the code rate, $M$ is the number of symbols in a block, $\rho$ is the fraction of useful resources utilized for data and $T_s$ is the symbol duration. 

\subsubsection{Related Work}

Design of physical layer processing tailored for sub-THz under the prominent hardware impairment of PN is reported in \cite{bicais2020digital}, introducing a PN-robust analytical constellation design named Polar-QAM. However, the PN-robustness is achieved at the expense of an increased peak-to-average power ratio (PAPR). An ML-driven constellation design constrained under a PAPR limit has been introduced in \cite{marasinghe2024constellation}. The system is modeled as a differentiable end-to-end system and optimized for binary cross-entropy between transmit bits and their soft estimates at the receiver using back-propagation facilitated by the ML framework. Extending this to single-carrier waveform optimization by incorporating the transmit and receiver filters under PN while delivering low PAPR under a given ACLR for a certain bandwidth is explored in \cite{marasinghe2024waveform}. Employing PN-aware demapping rules as well as an NN at the receiver as the de-mapper shows the potential of learned transceivers with increased PN-robustness while delivering low PAPR boosting BLER performance and spectral efficiency. Preliminary testing of the feasibility of the aforementioned ML-driven solutions in real sub-THz hardware is reported in \cite{nguyen2024subthz}, which demonstrates satisfactory error vector magnitude (EVM) and decoding performance with reduced PAPR, showcasing the practical usability of the learned solutions.

\subsection{Competing Networks in Accord}

\subsubsection{Motivation}

Wireless networking in unlicensed bands involves nearby networks (belonging to the same administrator or not) competing for the same spectrum. To ensure fair access, regulatory entities have established basic rules, including the use of the listen before talk (LBT) mechanism, and explicit limits on transmission power to reduce interference, as well as on the maximum allowed transmission durations. Once these requirements are met, wireless stations have significant flexibility in their configurations. They can select their operating channel, determine channel width, use single or multiple channels concurrently, adjust transmission power, and choose whether to deliver energy in all directions or focus it toward specific directions. This flexibility leads to a vast array of possible configurations and complex interactions between coexisting networks in frequency, time, and space. Those interactions may lead to a plethora of effects in the wireless domain, leading to poor performance if the parameters of each network are not carefully adjusted. Given the high complexity of the resource allocation problem in competing wireless networks---where the variability across deployments is huge---designing a one-fits-all solution becomes a challenging task. 

\subsubsection{Description}

To address the problems that arise from competition in coexisting wireless networks, solutions ranging from decentralized to centralized approaches can be considered:
\begin{itemize}
    \item \textbf{Decentralized/Distributed:} First, fully decentralized solutions include mechanisms that operate based on only local information and without any communication among competing devices. This is the case of the current channel access protocols in place, such as the DCF in Wi-Fi, which aims to mitigate concurrent channel accesses by multiple devices through randomness. In multi-BSS settings, decentralized solutions are optimal only if the optimization of local objectives leads to the optimization of the whole network (in game theory terms, a \textit{Pareto optimal} solution). However, this is typically not the case in real networks, where devices may aim at transmitting as much as possible and under the highest possible quality (e.g., using higher transmit power to achieve higher SNR), but the caused interference generates a negative impact on neighboring devices.
    \item \textbf{Cooperative:} Another type of solution is based on the existence of cooperation, or some sort of coordination, among competing devices. Accordingly, the involved devices are able to exchange information (e.g., configuration capabilities, performance experienced, reward signals) or commands, which are integrated into their configuring procedure and expected to achieve a better collective performance objective (e.g., by enabling new transmission possibilities for efficient spectrum usage through spatial reuse, coordinated beam-forming, or bandwidth sharing). Cooperative solutions, however, require additional capabilities to infer neighboring capabilities and goals, or even explicit inter-device communication (thus additional overheads). In both cases, cooperation may be unfeasible in current implementations, especially among independently operated wireless networks.
    \item \textbf{Centralized:} Finally, centralized solutions allow a network manager (or authorized entity) to decide the whole network configuration, thanks to a comprehensive view of the network. Centralized solutions can lead to optimal global performance while enforcing fairness, but at the same time require significant signaling and monitoring overheads and scale poorly due to the increased complexity associated with it (i.e., the configuration space increases exponentially with the number of devices). Moreover, centralized solutions require the agreement and onboarding of the different coexisting networks, which is today unfeasible in most of the current Wi-Fi deployments (e.g., residential Wi-Fi).
\end{itemize}

\begin{figure}[th!]
    \centering
    \centering
     \begin{subfigure}[b]{0.5\textwidth}
         \centering
         \includegraphics[width=\textwidth]{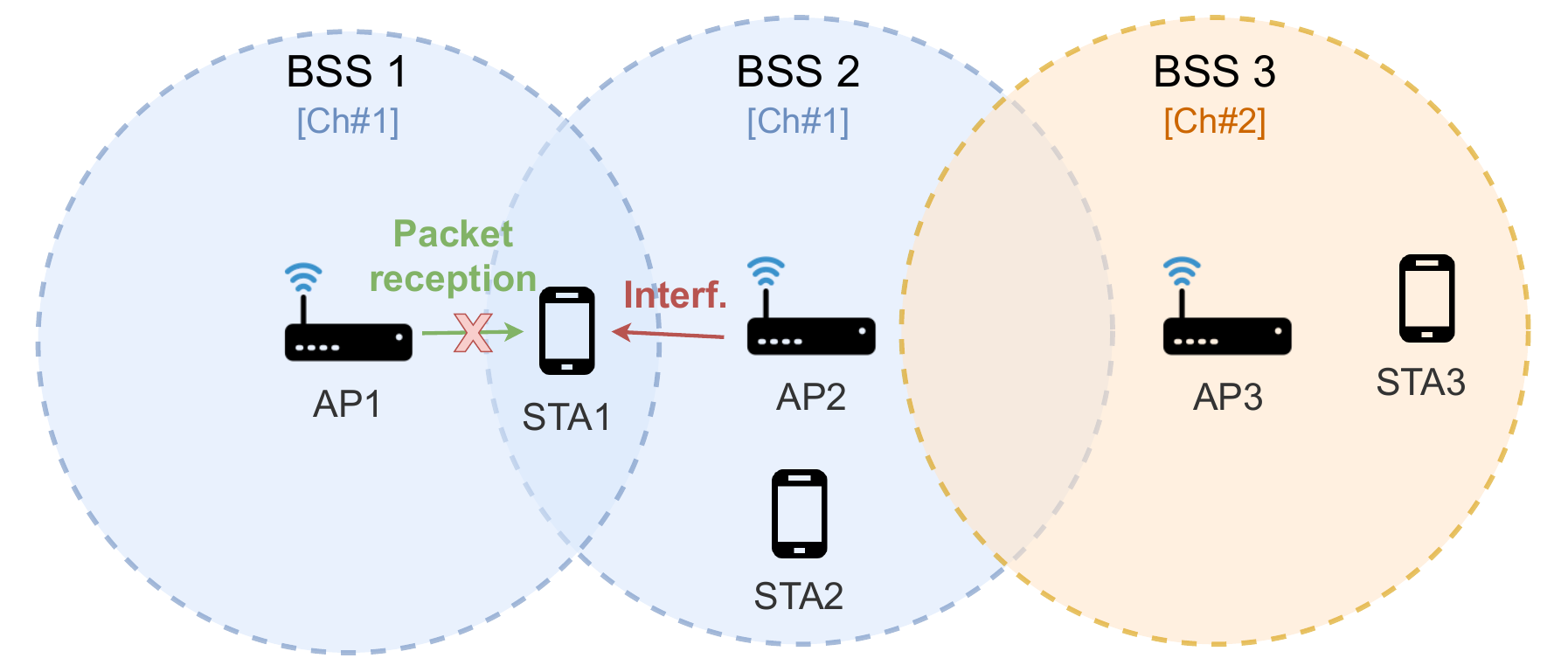}
         \caption{}
         \label{fig:CoexistingNetworks_a}
     \end{subfigure}
     \hfill
     \begin{subfigure}[b]{0.5\textwidth}
         \centering
         \includegraphics[width=\textwidth]{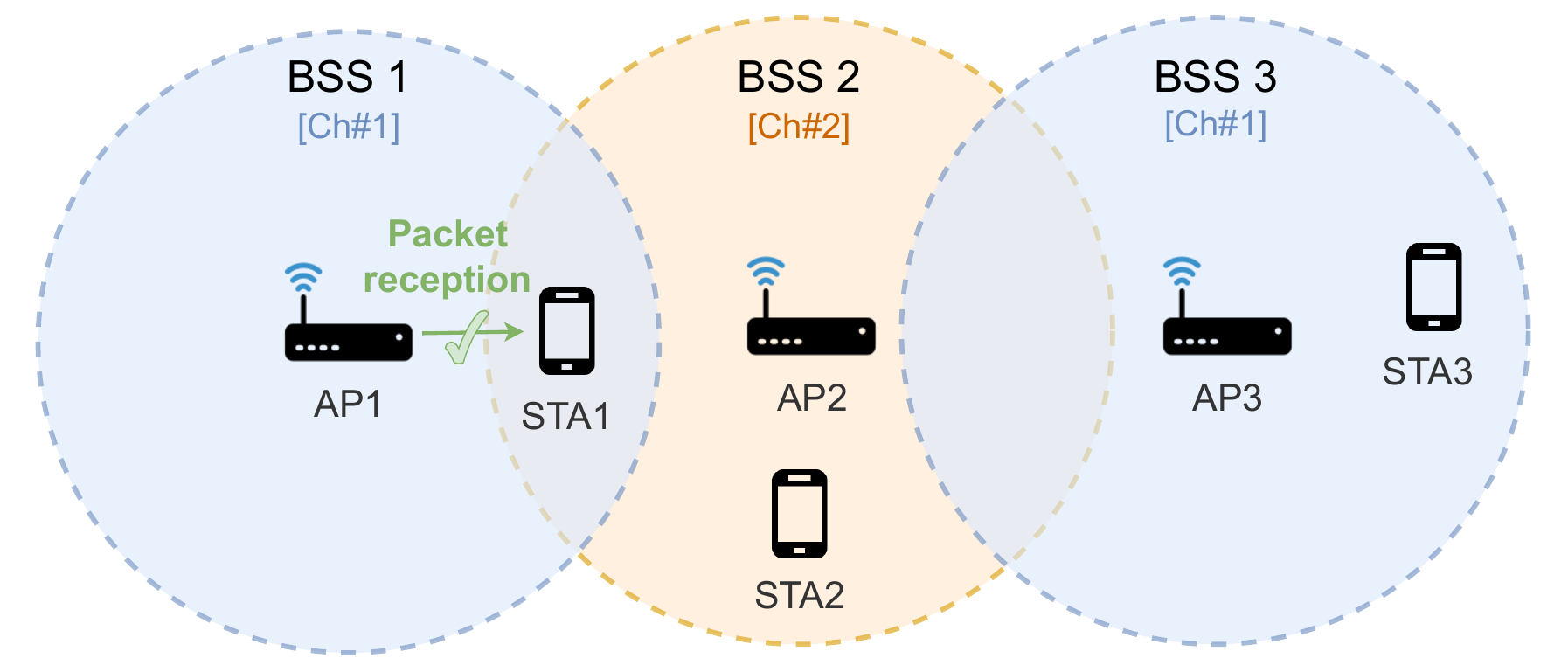}
         \caption{}
         \label{fig:CoexistingNetworks_b}
     \end{subfigure}
    \caption{The channel allocation problem in an example deployment. Default (static) configuration (a); dynamic (MLDR-based) configuration (b).}
    \label{fig:CoexistingNetworks}
\end{figure}

Our MLDR approach includes a coexistence management module that can address the trade-off between decentralization and centralization, thus dynamically determining the best approach in each particular case. While decentralized competition might be preferable in situations where a globally optimal solution is easy to achieve (e.g., in low-density deployments), in some other cases it might lead to poor performance. Likewise, cooperative and centralized solutions may remove some of the negative interactions in competing networks, but at the cost of additional overheads.
Figure~\ref{fig:CoexistingNetworks} illustrates a scenario with 3 competing networks. In particular, a default (static) solution might lead to suboptimal performance (Fig.~\ref{fig:CoexistingNetworks_a}), but the use of MLDRs can inter-network dependencies for good by learning from environmental interactions and dynamically adapting to changing conditions (Fig.~\ref{fig:CoexistingNetworks_b}). 

The MLDR approach proposed in this paper is based on an RL-oriented multi-agent setting that flexibly empowers the different players with the necessary capabilities, depending on the situation. The behavior of the agents, when attempting to learn a good-performing configuration, is driven by the observed reward, which in the wireless domain can take multiple forms, depending on the selected target. Relevant KPIs upon which agent rewards can be constructed include:
\begin{itemize}
    \item \textbf{Throughput ($\Gamma$):} The amount of MAC-layer data transmitted in bits per second (bps). The MAC throughput is computed from the Aggregated MAC Service Data Unit (A-MSDU) or Aggregated MAC Protocol Data Unit (A-MPDU) data successfully transmitted per second, and is given by
    \begin{equation}
        \Gamma = \frac{\text{L}_\text{Data}}{\text{T}_\text{Data} + \text{T}_\text{Overhead}}, 
    \end{equation}
    where $\text{L}_\text{Data}$ is the amount of data transmitted, $\text{T}_\text{Data}$ is the time it takes to transmit those data (which depends on the MCS selected according to channel quality), and $\text{T}_\text{Overhead}$ is the time spent with overheads (including control signals, contention time, collisions, etc.).
    \item \textbf{Delay ($l$):} The time it takes, in seconds (s), for a given data transmission to be successfully performed. The delay or latency might be affected by different procedures (e.g., queuing, processing, channel access), some of which might be considered to construct a reward. The delay can be computed as
    \begin{equation}
        l = T_2 - T_1, 
    \end{equation}
    where $T_1$ and $T_2$ are the initial and final timestamps under consideration, respectively (e.g., arrival of the data at the transmitter's queue and reception of the acknowledgment of its arrival from the receiver).
    \item \textbf{Spectral efficiency per area ($\eta$):} The amount of data transmitted per spectrum used in a given area, in bits per second per Hertz and cubic meter (bps/Hz/m$^3$).
    \begin{equation}
        \eta = \frac{\Gamma}{\text{BW} \times A},
    \end{equation}
    where $\Gamma$ is the throughput (see expression above), $\text{BW}$ is the bandwidth used, and $A$ is the area under consideration.
\end{itemize}

Based on the criteria above (and others), reward functions can be constructed, including individual performance metrics or a combination of them. Moreover, depending on the selected approach, rewards can be decentralized, shared, or centralized. In all cases, the definition of the reward must lead to a solution that satisfies application requirements (i.e., throughput, latency, and reliability). When cast into RL terms, an agent will attempt to minimize its regret by performing exploration. The regret is quantified as
\begin{equation}
    \upsilon = \sum_{i\in I} \phi^*_i - \phi_i,
\end{equation}
where $\phi^*_i$ is the optimal reward (achieved through the best possible action) in the control period $i\in I$ and $\phi_i$ is the actual reward that was experienced in that same period.

Quantifying the theoretical convergence time---the time it takes for an agent to learn an optimal policy (when possible)---of RL algorithms such as Q-learning (cf. ~\cite{even2003learning}) can however be very challenging in competing multi-agent setups. When it comes to cooperative or centralized approaches, better convergence guarantees can be provided at the expense of additional implementation cost~\cite{bistritz2020cooperative}. For that reason, it is critical that MLDRs are aware of the properties and implications of each approach (e.g., computational cost, convergence time, memory requirements, etc.), and use such knowledge as part of their configuration strategy. For instance, the tolerance of a given application to runtime exploration must be considered when selecting the best strategy to be implemented; while data transfer may tolerate exploratory actions with suboptimal performance, interactive video streaming cannot afford such disruptions.

\subsubsection{Related Work}

Multi-agent solutions for wireless networks have been significantly adopted for several years now~\cite{chen2021distributed, feriani2021single}. The multi-agent setting offers a good trade-off between complexity and performance, which suits well the typical high variability and partial observability of wireless systems.

A classic problem in which multi-agent solutions are used is channel allocation, given its orthogonality and the possibility for defining decentralized solutions with strong guarantees. The fact that the channel status can be represented as a binary variable (busy or idle) simplifies the problem and relaxes the uncertainty in wireless networks. In \cite{leith2006self}, a decentralized, lightweight learning algorithm was introduced to independently manage channel selection at the AP. The algorithm operates by taking simple actions---stay on the current channel or switch to a new one---, based on local measurements regarding previous successes/failures when accessing the channel. A similar approach was taken in~\cite{barrachina2021multi} for a more complex problem involving bandwidth selection (primary and secondary channels). To do that, the authors proposed a multi-agent MAB (MA-MAB) solution and compared multiple well-known algorithms. Interestingly, ~\cite{barrachina2021multi} reveals that, for that problem, simpler approaches like $\epsilon$-greedy perform better than others like upper confidence bound (UCB) or Thompson sampling (TS), which suggests that more refined parameter estimation methods are less effective in dynamic, chaotic multi-agent environments. More recently, with the advent of Wi-Fi 7 multi-link operation (MLO), the authors of~\cite{iturria2023channel} proposed a solution whereby each AP radio has an independent agent for channel selection, rather than a single agent managing all interfaces. These agents share their actions and observed rewards to build a joint Q-function (via transfer learning), reflecting a consensus that accelerates convergence. This multi-agent approach aligns with the objectives of the MLDR proposal.

When it comes to more complex problems, such as those involving power adaptation, the utility behind agents' rewards becomes harder to model, hence finding optimal solutions is not trivial due to the non-binary effects of interference. In~\cite{maghsudi2014joint}, joint channel selection and power adaptation mechanism based on multi-player MABs is proposed. In particular, this decentralized solution is proved to converge to a correlated equilibrium by relaxing the formulation of the reward and disregarding the interference coming from other devices. In~\cite{ wilhelmi2019collaborative, wilhelmi2019potential}, decentralized spatial reuse (integrating power adaptation with sensitivity adjustment) was addressed again as a multi-player MAB game. In that case, it was shown that, in certain scenarios, a selfish reward led to an optimal correlated equilibrium. However, this statement could not be generalized, provided that the competition unleashed among different networks could also lead to suboptimal results, thus being counterproductive in terms of performance.

To address that, cooperative approaches such as the one in~\cite{wilhelmi2024coordinated} have demonstrated the effectiveness of coordination among agents. In particular,~\cite{wilhelmi2024coordinated} explores the usage of several joint rewards (average, max-min, and proportional fairness), which are shown to outperform decentralized ones in performance and fairness. Another cooperative MAB approach for spatial reuse confirms the suitability of joint reward strategies~\cite{iturria2024cooperate}. Both~\cite{wilhelmi2024coordinated, iturria2024cooperate} require certain communication between agents (BSSs), so that reward signals can be exchanged. A step further is taken in~\cite{wojnar2024ieee, bardou2022inspire}, where semi-centralized solutions (e.g., where a controller determines groups of collaborating agents) are proposed for the same problem. 

Finally, it is worth reviewing other kinds of contributions in which decentralized methods can be improved by better processing locally available information. One example is~\cite{girmay2021machine}, which employs convolutional neural networks (CNNs) to estimate the saturation levels of Wi-Fi networks using inter-frame spacing statistics. Using this information, an agent is well-equipped to perform dynamic channel selection. Another relevant example is~\cite{shen2019graph}, which proposes the usage of graph neural networks (GNNs) to thoroughly process the channel information in wireless networks, so that efficient power adaptation can be performed. Overall, these communication-free methods empower decentralized approaches that work using local information only to the maximum extent. However, this comes at the expense of integrating more complex operations, often involving additional hardware and resources.

Taken together, these studies demonstrate the transformative potential of multi-agent RL in wireless networks. They underscore the necessity of balancing simplicity and sophistication in algorithm design, fostering structured agent interactions, and leveraging shared insights to enhance network performance across diverse scenarios. These principles form the cornerstone of emerging intelligent, adaptive communication systems.



\section{Channel Access in Bespoke Industrial Networks -- Results}
\label{sec:results}

To provide an example of an MLDR use case realization through AI/ML and highlight the arising flexibility of SpecNets, we focus on optimizing bespoke industrial networks (Section~\ref{sec:bespoke}). 
The problem that we address is being able to fulfill operator requirements, given as points on the ``SpecNets triangle'' (Figure~\ref{fig:triangle}). Specifically, we turn to MABs, which can learn online without \textit{apriori} knowledge
and dynamically configure network parameters to meet the varying performance goals.
In the following, we focus on the three scenarios listed in Table~\ref{tab:services} and explain the implementation, scenarios, and achieved results.

\subsection{Implementation}
\begin{figure}[t!]
    \centering
    \adjincludegraphics[width=0.95\columnwidth,trim={0 {0.25\height} 0 {0.2\height}},clip]{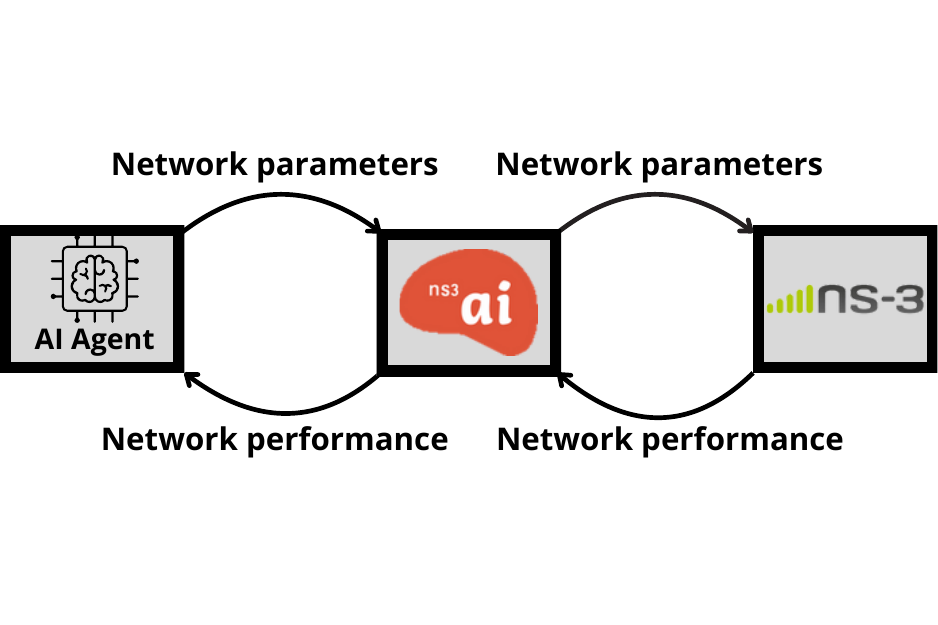}
    \caption{Integration of the MAB agent into the ns-3 simulator established with shared memory through the ns3-ai library. The simulator provides the agent with network performance metrics, while the agent returns the updated network configuration at each \textit{control period}.}
    \label{fig:ns3_ai}
\end{figure}

The implementation of a task-oriented network scenario involves using a network simulation environment driven by a MAB agent. 
The environment used in this work is ns-3.42\footnote{\url{https://www.nsnam.org}} as it allows the simulation of 802.11 networks with flexible network configurations. Furthermore, we use ns3-ai~\cite{ns3ai} to create a communication interface between the simulator and the MAB agent (Figure~\ref{fig:ns3_ai}).

The agent is implemented using the Reinforced-lib library~\cite{wojnar2024reinforced}. We select Thompson sampling as the action-selection strategy used by the agent due to its simplicity and mathematical foundation, which ensures asymptotic regret in stationary cases~\cite{slivkins2019introduction}. 
The agent configures the following channel access rules and parameters (constant in each control period):
\begin{itemize}
    \item the contention window (CW) value, limited to discrete values of $2^{n+4}-1$, where $n \in {0, 1, \ldots, 6}$,
    \item enabling or disabling A-MPDU frame aggregation,
    \item enabling or disabling the RTS/CTS frame protection mechanism.
\end{itemize}
The same configuration is applied to all stations controlled by the AP. More freedom (independent, per-station configurations) would lead to even better results at the expense of complicating the problem significantly.
A repository containing the source code for the simulation model, the agent, and the scripts to reproduce the experiments is available as open source.\footnote{\url{https://github.com/mldr-devs/mldr}}

\subsection{Reward Design}
To optimize the system for high throughput, low latency, and fair resource allocation among stations, we define a multi-objective reward function as a weighted sum of normalized metrics. The reward combines three components (Figure~\ref{fig:3d_triangle}): overall fairness $f$ (measured using Jain's fairness index calculated over per-station throughput), aggregate throughput $t$, and mean latency $l$. Each of the objectives is weighted by their respective importance $w_i$, where $i \in \{f, t, l\}$ and $\sum_i w_i = 1$. The fairness reward $r_f$ is modeled as $r_f = 1 - \alpha (1 - f)$, where $\alpha$ adjusts the sensitivity to fairness. This form encourages even resource allocation by increasing the reward as fairness increases. The throughput reward $r_t$ is represented as $r_t = t / t_{\text{max}}$, where $t_{\text{max}}$ is the maximum data rate achievable, and $t$ is the actual throughput obtained at a given iteration. The latency reward $r_l$ is defined as $r_l = \beta / l$, with $\beta$ an adjusting latency weight. By inversely rewarding latency, $r_l$ prioritizes low-latency actions. Each component $r_i$ is clipped between 0 and 1 to prevent any single metric from dominating. The final reward is thus expressed as:
\begin{equation}
    R = \sum_i w_i \cdot \min(1, \max(0, r_i)).
\end{equation}
This is a form linear
scalarizatio~\cite{drugan2013moma} and weights $w_i$ can be easily configured using\eqref{eq:spec_eff}  

\subsection{Scenarios}
\begin{figure}
     \centering
     \begin{subfigure}[b]{0.22\textwidth}
         \centering
         \includegraphics[width=\textwidth]{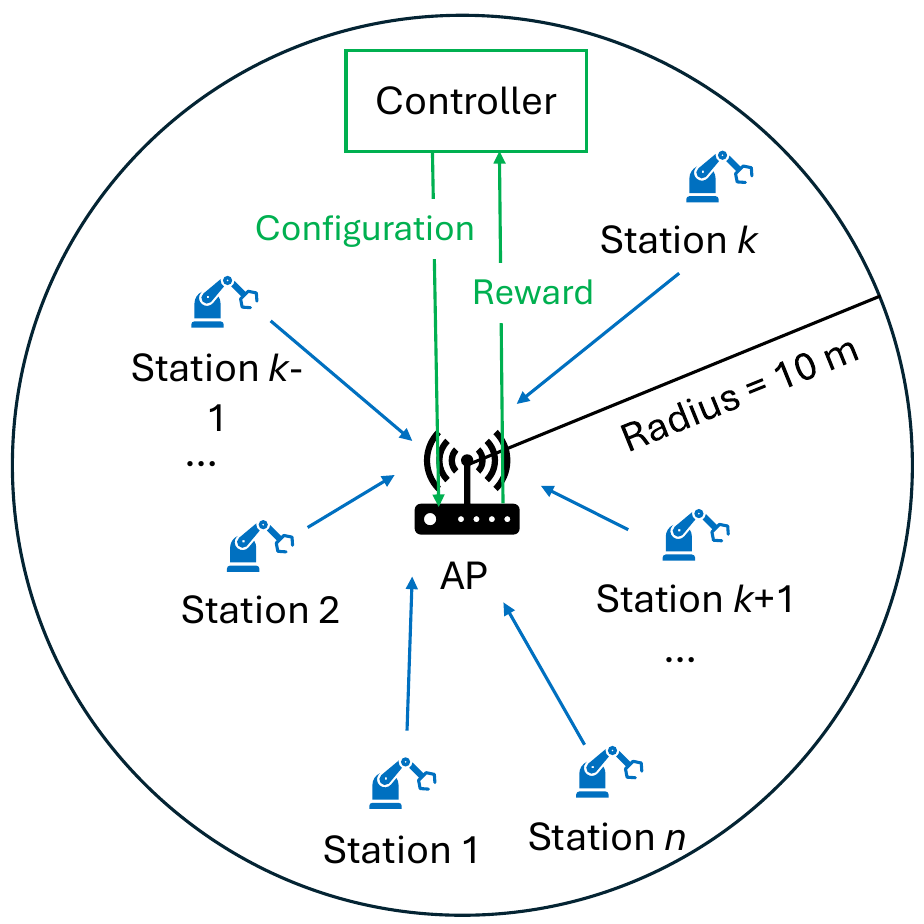}
         \caption{}
         \label{fig:sim-scenarios-1}
     \end{subfigure}
     \begin{subfigure}[b]{0.25\textwidth}
         \centering
         \includegraphics[width=\textwidth]{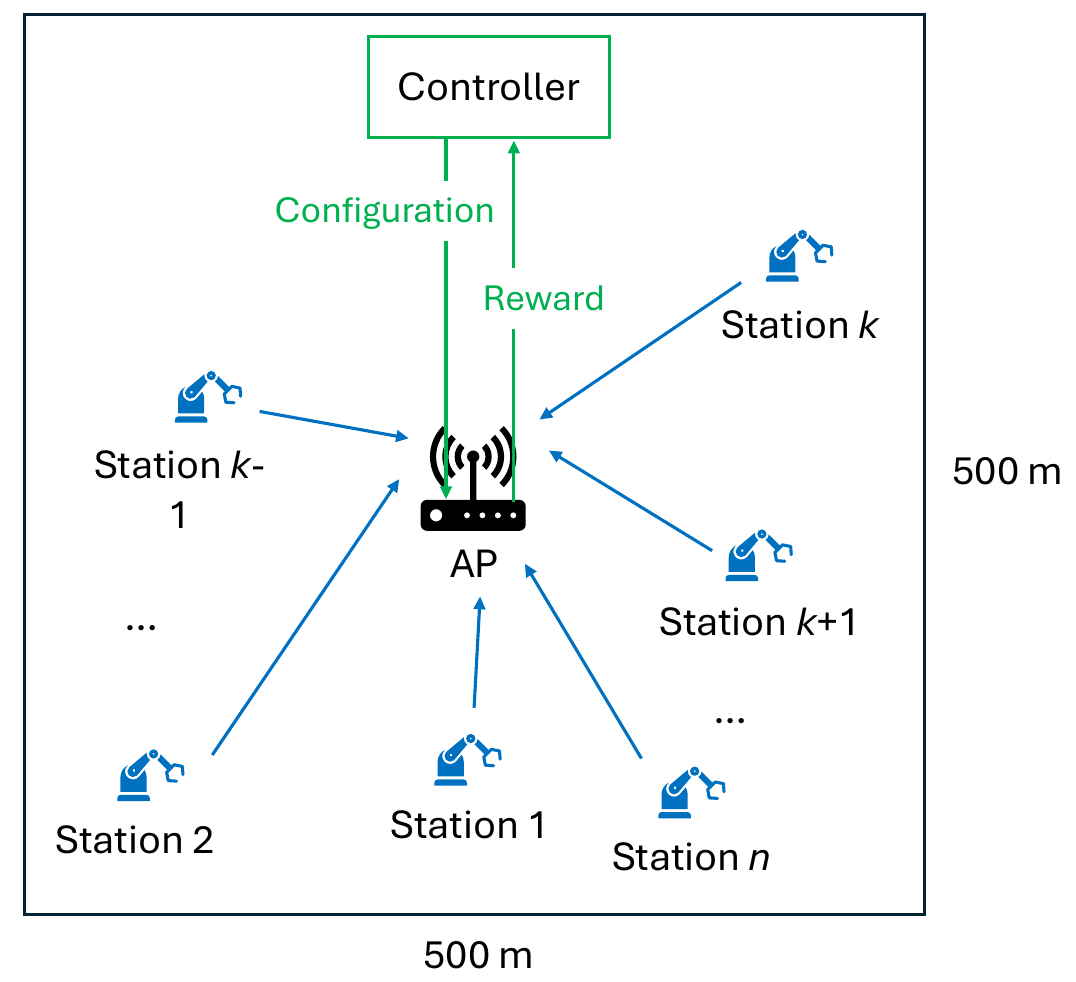}
         \caption{}
         \label{fig:sim-scenarios-2}
     \end{subfigure}
        \caption{Simulation scenarios high throughput and low latency (a); massive communications (b).}
        \label{fig:simulation-scenarios}
\end{figure}

Our study examines three prominent use-case scenarios with extreme requirements in bespoke industrial networks (high throughput, low latency, and massive communication), where each scenario assumes the maximization of one of the parameters at the expense of others. The scenarios consider topologies tailored to their specific needs. Both low latency and high throughput scenarios operate as traditional local wireless networks composed of one access point (AP) and its associated stations, which are stationary and placed randomly within a \SI{10}{\meter} radius (Fig.~\ref{fig:sim-scenarios-1}). Each station generates identical constant bit rate uplink traffic (Table~\ref{tab:parameters_scenario}). To maximize performance, we set a constant PHY transmission rate (MCS~11). Meanwhile, in the massive communication scenario, stations are randomly distributed within a square area of \SI{500}{\meter} $\times$ \SI{500}{\meter} with the AP in the middle (Fig.~\ref{fig:sim-scenarios-2}). Due to the high range required, we set the constant PHY transmission rate to MCS~0. Each station transmits a small, \SI{256}{\byte} packet to the AP every \SI{50}{\milli\second}, mirroring an industrial sensor network that sends measurement data.

Tables~\ref{tab:parameters_general} and \ref{tab:parameters_scenario} present the general and scenario-specific parameters, respectively. As MAB agents require some time to reach stability, by default, their performance is evaluated after an initial warm-up phase lasting up to 200 \textit{control periods}, unless explicitly stated. In practice, this warm-up phase is often much shorter and is determined by a straightforward heuristic that indicates convergence. The warm-up phase ends if the agent selects the same network configuration at least 90\% of the time during the last 20 \textit{control periods}.

\begin{table}[t!]
    \centering
    \caption{General simulation parameters and default IEEE 802.11 settings.}
    \label{tab:parameters_general}
    \begin{tabular}{@{}ll@{}}
        \toprule
        \textbf{Simulation parameter}     & \textbf{Value}   \\ \midrule
        Band                          & \SI{5}{\giga\hertz}    \\
        PHY/MAC                          & IEEE 802.11ax (no OFDMA)    \\
        Spatial   streams          & 1, SISO  \\
        Channel model                     & Log-distance    \\
        Channel width                     & 20 MHz          \\  
        CWmin/CWmax              & 15/1023 \\
        A-MPDU size              & \SI{64}{\kilo\byte} \\
        RTS/CTS & Disabled \\
        Control period                    & 0.5 s             \\ 
        Number of independent runs        & 10                \\ \bottomrule
    \end{tabular}
\end{table}

\begin{table*}[t!]
    \centering
    \caption{Scenario-specific simulation parameters.}
    \label{tab:parameters_scenario}    
    \begin{tabular}{@{}lllll@{}}
        \toprule
        \textbf{Scenario}    & \textbf{No. of stations} & \textbf{Packet size (payload)} & \textbf{Load per station} & \textbf{MCS}   \\ \midrule
        High throughput          & 5 -- 50  & 1500 B & \SI{120}{\mega\bit\per\second} (full buffer) & 11\\            
        Low latency              & 4        & 1500 B & \SI{20}{\mega\bit\per\second} & 11 \\
        Massive communication    & 250      & 256 B  & \SI{46}{\kilo\bit\per\second} & 0  \\    \bottomrule
    \end{tabular}
\end{table*}


\begin{figure*}[ht!]
    \centering
    \subfloat[Evolution of instantaneous aggregate network throughput]{
        \includegraphics[width=\textwidth]{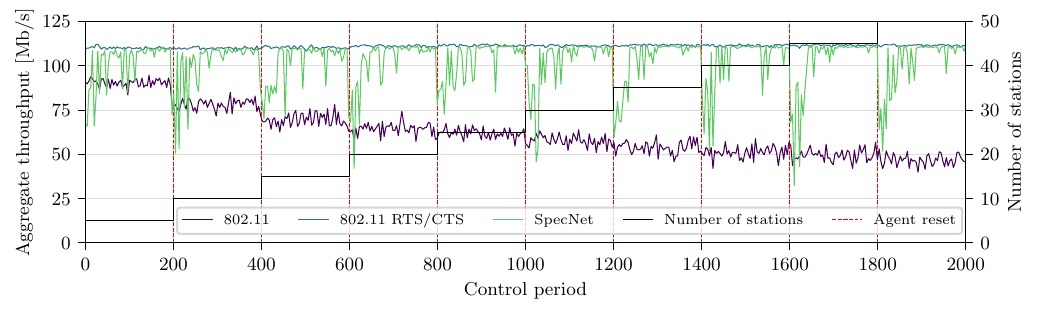}	
        \label{fig:dynamicTHR}
    }   
    \\
    \subfloat[Evolution of instantaneous fairness]{
        \includegraphics[width=\textwidth]{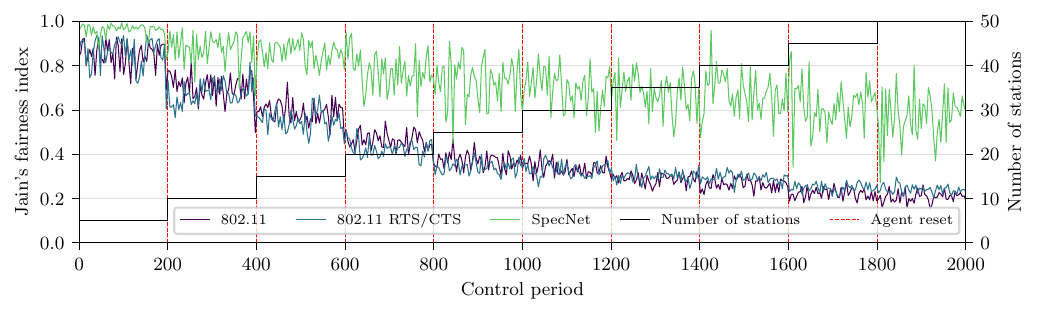}	
        \label{fig:dynamicFAIRNESS}
    }   
    \\
    \caption{Performance evolution in the dynamic high throughput scenario.}
    \label{fig:throughput_plots}
\end{figure*}


\begin{figure}[ht!]
    \centering
    \subfloat[Training process.]{
        \includegraphics[width=0.45\textwidth]{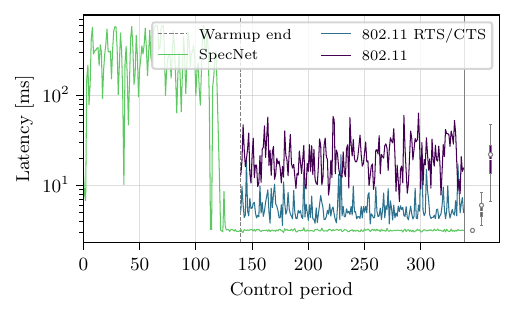}	
        \label{fig:latency_train_test}
    }   
    \\
    \subfloat[CDF of per-packet latency.]{
        \includegraphics[width=0.45\textwidth]{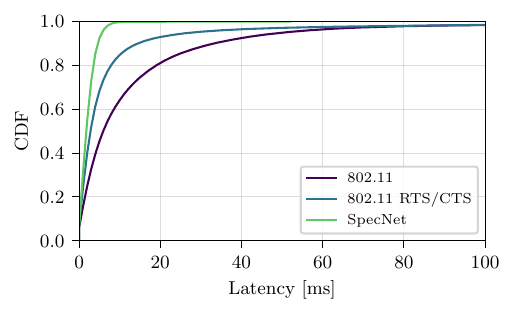}	
        \label{fig:latency_cdf}
    }   
    \caption{Results for the low latency scenario.}
    \label{fig:latency_plots}
\end{figure}

\subsection{Results}

In the following, we present the results obtained in each scenario by the different studied mechanisms. We use IEEE 802.11 in its default configuration of Table~\ref{tab:parameters_general} (referred to as \emph{802.11}) as a baseline to evaluate the performance of SpecNets.
We also adopt the same 802.11 configuration but with RTS/CTS enabled, which we name \emph{802.11 RTS/CTS}.
For the SpecNet configuration, note that it may vary across the three evaluated scenarios, given the dynamism of the solution. 

\subsubsection{High Throughput Scenario}
Figure~\ref{fig:throughput_plots} presents the results of a dynamic scenario focused on throughput optimization. Every \SI{100}{\second}, 5 additional stations are added to the network and the MAB agent is restarted to its initial state. As can be seen in Figure~\ref{fig:dynamicTHR}, after approximately 100 control periods the network reaches the optimal configuration. Convergence is also achieved after every consecutive change in the network. The agent learns to obtain a high reward by enabling RTS/CTS, aggregation, and adjusting the CW setting to the degree of congestion. Similar performance can be achieved using 802.11 but only with RTS/CTS enabled. 
Disabling this frame protection mechanism lowers the aggregate network throughput as the network size increases and packet collisions have an increasingly negative impact on performance.

Even though our agent does not find a solution which would surpass 802.11 performance with RTS/CTS enabled (which is already close the the upper bound), the SpectNet agent finds a solution which is much better in terms of instantaneous fairness as shown in Figure~\ref{fig:dynamicFAIRNESS}.
The IEEE 802.11 channel access function is known for its short term unfairness \cite{miskowicz2021unfairness}, but the agent solves this by optimizing the CW value so that instantaneous fairness is higher without any loss in aggregate throughput.

\subsubsection{Low Latency Scenario}
In the second studied scenario, the SpecNet agent focuses on achieving low end-to-end latency.
Figure~\ref{fig:latency_train_test} shows the temporary performance achieved by the agent, including an initial transitory phase (with intense exploration) and a steady state phase (where the learned distributions are exploited). 
At first, the network behavior is chaotic as the agent tries different configuration.
After about 140 control periods, the agent stabilizes after finding the optimal configuration. 
Compared to the default 802.11 configuration, the SpecNet-controlled network shows reduced average latency and variance, making its performance more predictable. Figure~\ref{fig:latency_cdf} illustrates the combined steady-state results of every run in the low latency scenario as a cumulative distribution function (CDF). 
Notably, the agent typically follows a similar network configuration (i.e., RTS/CTS, frame aggregation enabled, low CW) as in the high-throughput scenario. However, the difference between the performance of 802.11 with RTS/CTS and SpecNet  is clearly visible on account of two factors: (a) RTS/CTS improves latency by minimizing airtime wasted on collisions, (b) optimizing CW finds a balance between decreasing channel access times (lower CW) and minimizing collisions (higher CW).

\subsubsection{Massive Communication Scenario}
\begin{figure}[!t]
\centering
\includegraphics[width=0.95\columnwidth]{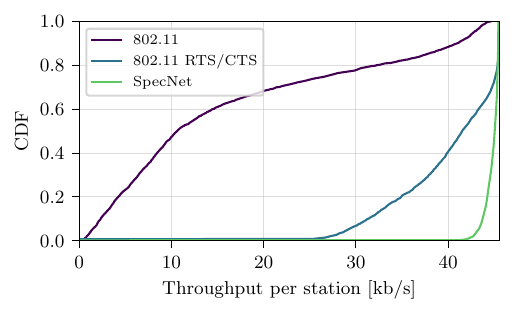}
    \caption{Throughput distribution in the massive communication scenario over multiple iterations.}
    \label{fig:massive_cdf}
\end{figure}

In the massive communication scenario, our goal is to ensure that as many devices as possible can access the channel.
Therefore, the SpecNet agent optimizes for instantaneous fairness.
The results in Figure~\ref{fig:massive_cdf} as a CDF of throughput achieved by each station clearly show the superiority of the proposed solution.
The SpecNet agent optimizes network performance with the following configuration: enabling RTS/CTS reduces time spent in collisions, enabling frame aggregation reduces the number of required channel access attempts, and using a higher CW value reduces the number of collisions.
With this configuration, the SpecNet agent outperforms both variants of 802.11 with nearly all of the 250 stations achieving the required offered load.

\subsubsection{Summary} 
Finally, Table~\ref{tab:config_performance_alt} presents the aggregate results of all 
scenarios in different static network configurations, compared to the configuration selected by the agent. The table contains the average metrics of 10 independent runs along with their standard deviation. This comparison helps assess how different network parameters influence performance.
We see the most efficient configuration provides both the highest throughput and lowest latency.
Such a situation may occur even if scenario settings and optimization targets are different, which highlights the robustness of our approach.
Therefore, this small but illustrative example confirms that a single SpecNet agents can adapt to various operator requirements in bespoke industrial networks.

\begin{table*}[t!]
    \centering
    \caption{Aggregate network performance for each scenario depending on parameter settings. 
    The results in bold indicate the best result for  fixed settings (top) and for the MAB agent (bottom).}
    \label{tab:config_performance_alt}
    \begin{tabular}{cccc|ccc} 
        \toprule
        \textbf{Agent} & \textbf{RTS/CTS} & \makecell{\textbf{Frame}\\\textbf{aggregation}} & \textbf{CW} & \makecell{\textbf{High throughput}\\ \textbf{[\SI{}{\mega\bit\per\second}]}} & \makecell{\textbf{Low latency}\\ \textbf{(90th percentile of delay) [\SI{}{\milli\second}]}} & \makecell{\textbf{Massive communication}\\ \textbf{(Jain's fairness index)}} \\ \midrule
        Fixed & \checkmark & \checkmark & Lowest (15)       & \textbf{110.3 $\pm$ 1.1} & \textbf{5.0 $\pm$ 0.7}  & 0.40 $\pm$ 0.04          \\ 
        Fixed & \checkmark & \checkmark & Highest (1023)       & 103.7 $\pm$ 0.9          & 26.4 $\pm$ 2.1          & \textbf{1.00 $\pm$ 0.00} \\ 
        Fixed & \checkmark & $\times$   & Lowest (15)        & 32.2 $\pm$ 0.8           & 616.8 $\pm$ 7.5         & 0.07 $\pm$ 0.01          \\ 
        Fixed & \checkmark & $\times$   & Highest (1023)       & 17.0 $\pm$ 0.0           & 699.7 $\pm$ 12.9        & 0.88 $\pm$ 0.01          \\ 
        Fixed & $\times$   & \checkmark & Lowest (15)        & 53.5 $\pm$ 3.4           & 56.6 $\pm$ 35.9         & 0.02 $\pm$ 0.00          \\ 
        Fixed & $\times$   & \checkmark & Highest (1023)       & 103.9 $\pm$ 0.9          & 25.0 $\pm$ 2.0          & 0.83 $\pm$ 0.01          \\ 
        Fixed & $\times$   & $\times$   & Lowest (15)        & 22.1 $\pm$ 0.5           & 606.3 $\pm$ 1.9         & 0.07 $\pm$ 0.01          \\ 
        Fixed & $\times$   & $\times$   & Highest (1023)       & 19.2 $\pm$ 0.0           & 702.6 $\pm$ 13.2        & 0.61 $\pm$ 0.02          \\ \midrule
        MAB   & Auto       & Auto       & Auto          & \textbf{108.9 $\pm$ 1.1} & \textbf{5.4 $\pm$ 0.8}  & \textbf{1.00 $\pm$ 0.00} \\ \bottomrule
    \end{tabular}
\end{table*}


\section{Conclusions}
\label{sec:conclusions}
Advances on AI/ML will enable the development of autonomous SpecNets, able to not only self-configure to handle some application and scenario requirements, but also to natively create new protocols to communicate that are tailored made to the problem to solve.

\section*{Acknowledgment}

This paper is supported by the CHIST-ERA Wireless AI 2022 call MLDR project (ANR-23-CHR4-0005), partially funded by AEI and NCN under projects PCI2023-145958-2 and 2023/05/Y/ST7/00004, respectively. The work of F. Wilhelmi and B. Bellalta is also partially supported by Wi-XR PID2021-123995NB-I00 (MCIU/AEI/FEDER,UE), and by MCIN/AEI under the Maria de Maeztu Units of Excellence Programme (CEX2021-001195-M). For the purpose of Open Access, the authors have applied a CC-BY public copyright licence to any Author Accepted Manuscript (AAM) version arising from this submission.



\ifCLASSOPTIONcaptionsoff
  \newpage
\fi



\bibliographystyle{IEEEtran}
\bibliography{bibliography}
%


%








\end{document}